\documentclass[a4paper,11pt]{article}
\usepackage{pos}
\usepackage{xspace}

\title{Machine-learning based particle-flow algorithm in CMS}

\author*{Farouk Mokhtar}
\onbehalf{on behalf of the CMS Collaboration}

\affiliation[]{Department of Physics \\
University of California San Diego \\
9500 Gilman Drive \\
La Jolla, CA 92093
}

\emailAdd{fmokhtar@ucsd.edu}

\abstract{
The particle-flow (PF) algorithm provides a global event description by reconstructing final-state particles and is central to event reconstruction in CMS.
Recently, end-to-end machine learning (ML) approaches have been proposed to directly optimize physical quantities of interest and to leverage heterogeneous computing architectures.
One such approach, machine-learned particle flow (MLPF), uses a transformer model to infer particles directly from tracks and clusters in a single pass.
We present recent CMS developments in MLPF, including training datasets, model architecture, reconstruction metrics, and integration with offline reconstruction software. 
}

\FullConference{ European Physical Society Conference on High Energy Physics (EPS-HEP 2025)\\
7-11 July 2025\\
Marseille\\}

\newcommand{\met}{\ptmiss}

\newcommand{\qcd}{QCD\xspace}

\newcommand{\ONNX} {{\textsc{ONNX}}\xspace}
\newcommand{\ONNXRUNTIME} {{\textsc{onnxruntime}}\xspace}

\newcommand{\akfourchs}[1]{AK4-CHS\xspace}
\newcommand{\akfourpuppi}[1]{AK4-PUPPI\xspace}

\newcommand{\ptmomentum}{\ensuremath{p_{\mathrm{T}}}\xspace}

\newcommand{\CMSSW} {{\textsc{CMSSW}}\xspace}
\newcommand{\GEANTfour} {{\textsc{Geant4}}\xspace}
\newcommand{\PYTHIA} {{\textsc{pythia}}\xspace}

\newcommand{\TeV}{\ensuremath{\,\text{Te\hspace{-.08em}V}}\xspace}
\newcommand{\ttbar}{\ensuremath{\mathrm{t}\overline{\mathrm{t}}}\xspace}
\newcommand{\kt}{\ensuremath{k_{\mathrm{T}}}\xspace}
\newcommand{\ptmiss}{\ensuremath{\ptmomentum^\text{miss}}\xspace}
\newcommand{\ptvecmiss}{\ensuremath{{\vec p}_{\mathrm{T}}^{\kern1pt\text{miss}}}\xspace}

\newcommand{\pt}{\ensuremath{p_{\mathrm{T}}}\xspace}

\begin{document}
\maketitle

\section{Introduction}

 The CMS particle-flow (PF) algorithm~\cite{CMS:2017yfk} uses rule-based methods, such as proximity-based linking---associating tracks and calorimeter clusters---to reconstruct a global, particle-level view of each event.
In contrast, machine-learned particle-flow (MLPF) uses transformer models trained on simulation to exploit low-level features of particle interactions with the detector that may not be immediately obvious from a first principles approach based on feature engineering.
In these proceedings, we present an MLPF implementation integrated within the CMS software framework (\CMSSW), trained on Monte Carlo (MC) simulation samples with pileup (PU) and validated both in simulation and on proton-proton collisions data collected during Run 3 (2022 -- 2026) by the CMS experiment~\cite{Chatrchyan:2008zzk,CMS:2023gfb}.

\section{Datasets and simulation samples}

The CMS detector response is simulated using \GEANTfour (v11.2.2)~\cite{GEANT42003,GEANT42006,GEANT42016}.
Parton showering, fragmentation, and hadronization are modeled with \PYTHIA (v8.311)~\cite{Bierlich:2022pfr}, using the CP5 underlying event tune~\cite{Sirunyan:2019dfx}.
Simulated MC samples for training and validation are produced at a center-of-mass energy of 14\TeV, under Run 3 (2023) detector and accelerator conditions, including \ttbar, QCD multijet, and all-hadronic \(\mathrm{Z} \to \tau\tau\) processes.
For samples with PU, we apply a flat profile with an average of 55--75 interactions per bunch crossing.
For validation on data, we use a small subset of CMS Run 3 data passing the dijet triggers, collected at a center-of-mass energy of 13.6\TeV.

\section{MLPF algorithm}

\subsection{Simulation-based target}

MLPF is trained to reconstruct \textit{target} particles that leave hits in the detector either directly or through their decay products, and is validated against \PYTHIA \textit{truth} particles.
Due to the finite granularity and thresholds in the detector simulation and reconstruction chain, there is an unavoidable smearing between the truth particles and the target particles.
For example, if a particle is below the energy threshold to leave hits in the \GEANTfour simulation, or if it does not produce any hits that survive upstream reconstruction (e.g., track or cluster reconstruction), it is excluded from the target and thus cannot be reconstructed.

\begin{figure}[ht]
    \centering
    \includegraphics[width=0.32\linewidth]{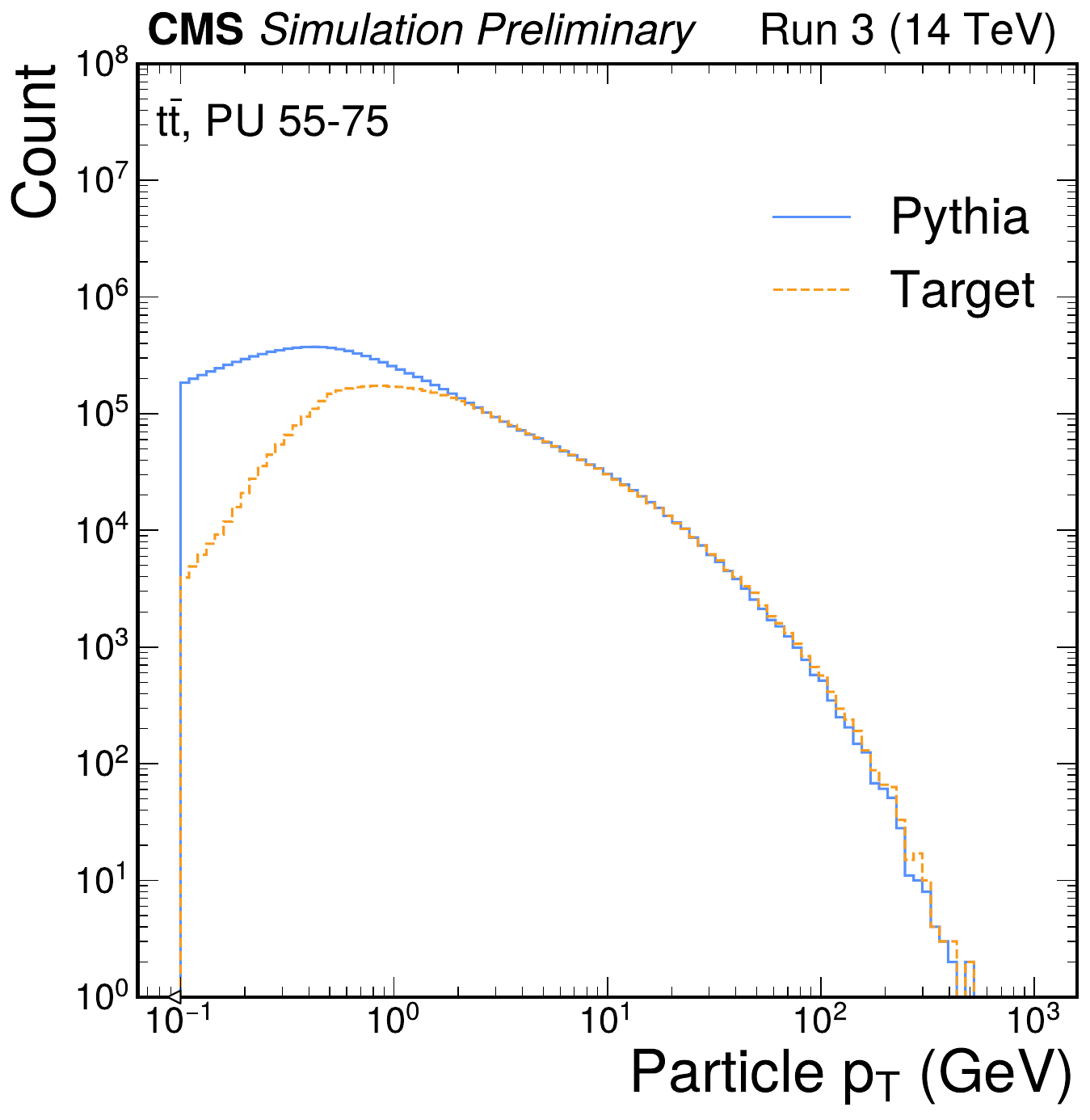}
    \includegraphics[width=0.32\linewidth]{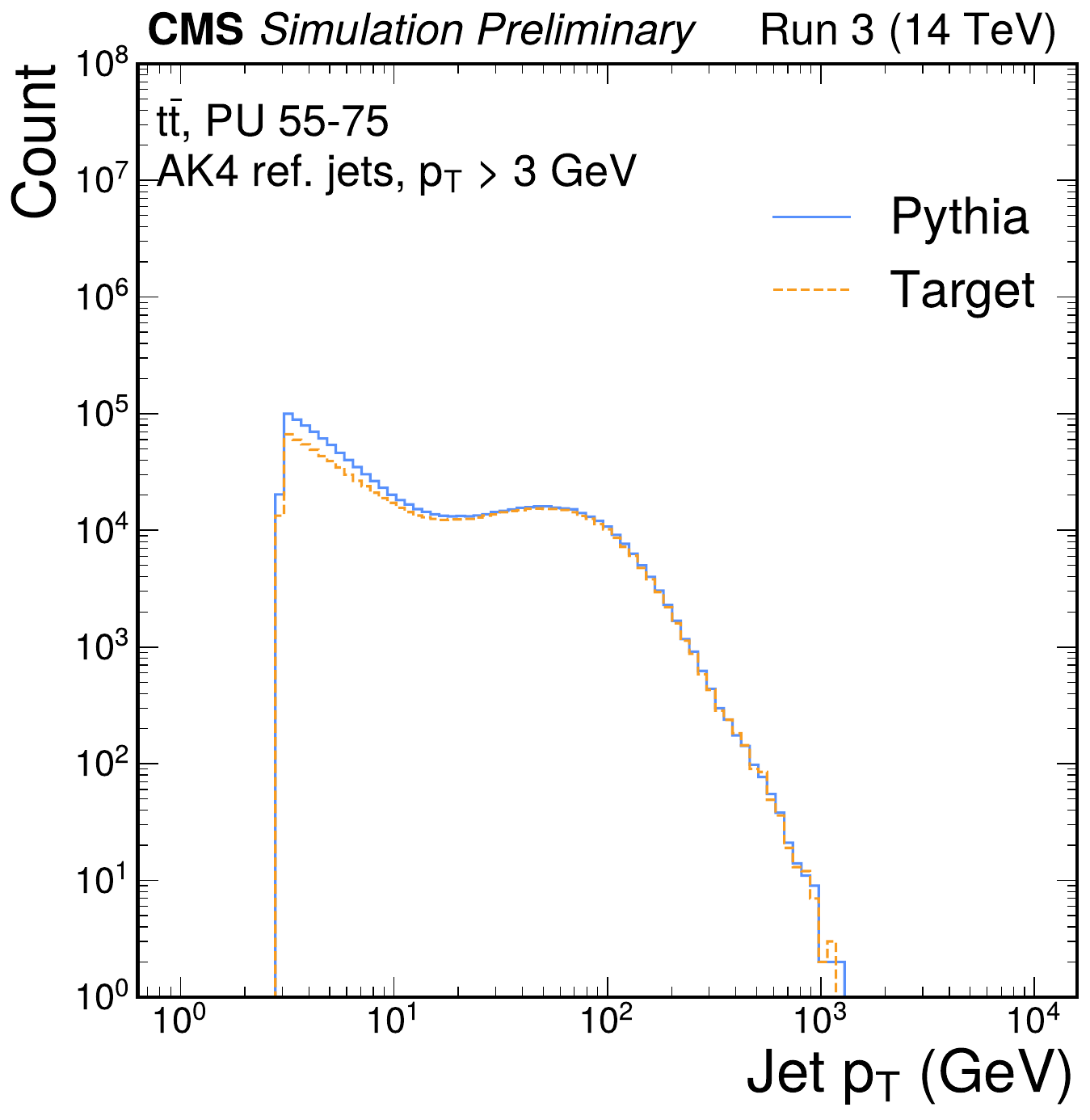}
    \includegraphics[width=0.32\linewidth]{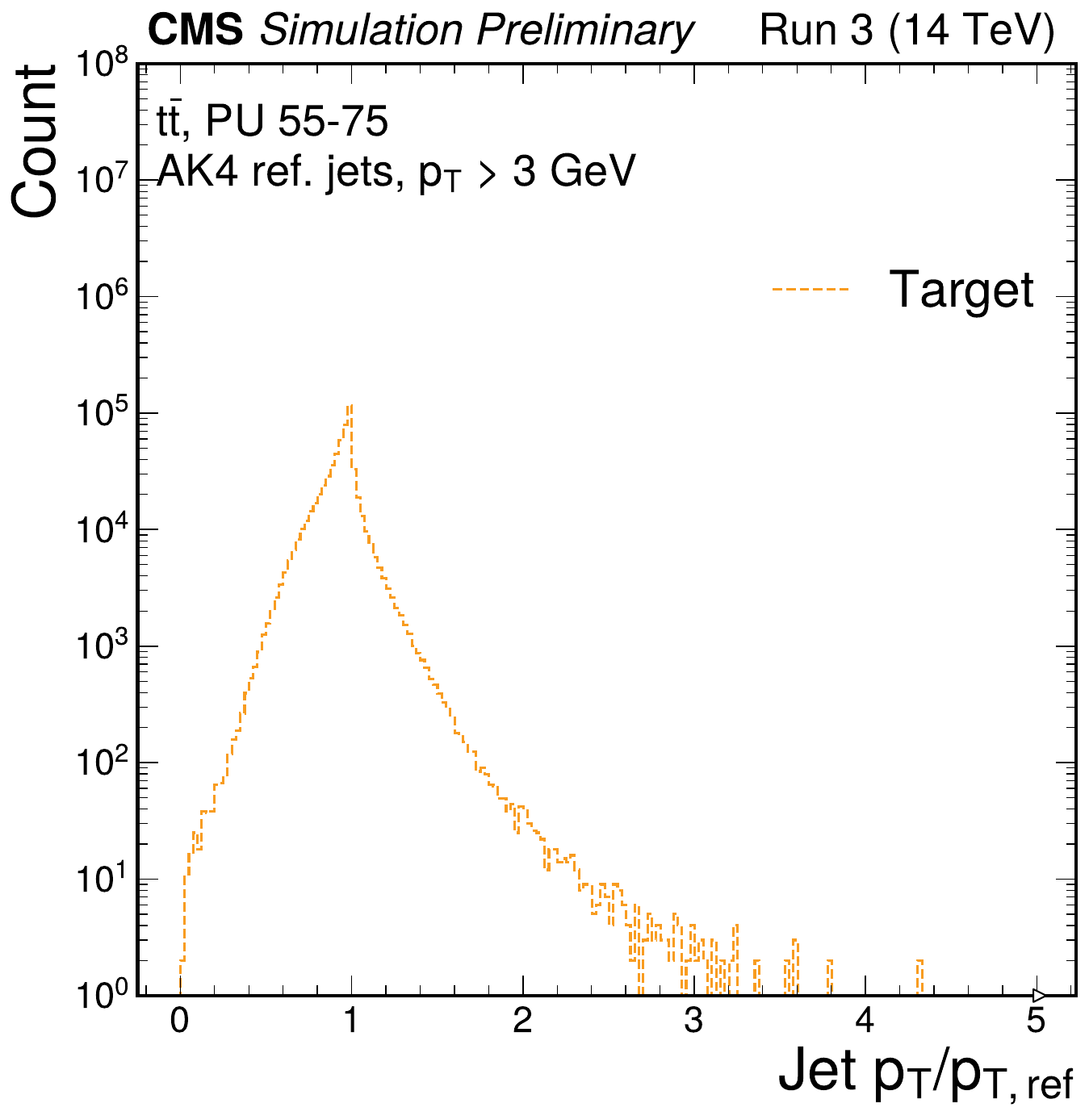}
    \caption{Validation of the MLPF target using \ttbar simulation with PU.
    Left: \pt distribution of target particles, and \PYTHIA truth particles.
    Center: target jet \pt distribution and \PYTHIA truth jets.
    Right: jet response vs. \PYTHIA truth jets.
    The target closely matches PU-subtracted \PYTHIA truth.
    }
    \label{fig:target_validation}
\end{figure}

In addition, for samples generated with PU, there is an additional mismatch between the MLPF target and the truth: while the truth set contains only hard-scatter particles, the target set includes particles from both the hard interaction and PU.
To handle this, we record the PU energy fraction associated with each target particle.
Additionally, we define a PU-masked target, which includes only the subset of target particles associated with the hard interaction, enabling validation and performance studies without PU contamination.
We validate the MLPF target definition using inclusive \ttbar samples with PU, confirming that the target aligns well with pileup-subtracted \PYTHIA truth particles, as shown in Fig.~\ref{fig:target_validation}.

\subsection{MLPF architecture and loss function}

The model architecture follows the original MLPF approach~\cite{Pata:2021oez,Pata:2022wam,Mokhtar:2023fzl}, replacing GNN layers with transformers accelerated with FlashAttention~\cite{dao2022flashattention} for improved scalability and performance.
The model takes as input a set of reconstructed tracks and calorimeter clusters, each represented by a feature vector $x_i$, collectively forming the set of PF elements $X$.
The input features for each track and cluster are described in Section~\ref{sec:inp}.
The model outputs reconstructed particle candidates $\hat{y}_i \in \hat{Y}$, trained to match a set of simulation-based target particles $Y$.
Each target particle $y_i$ is characterized by three quantities: (1) its particle identification (PID), (2) a binary label $\alpha_i$ indicating whether it originates from pileup, and (3) its four-momentum, parameterized as $(\pt, \eta, \sin{\phi}, \cos{\phi}, E)$.

Since the input and output sets have different cardinalities, we associate each target particle with a single unique input element, designated as the \textit{primary element}, to define a per-particle loss function.
This allows reconstruction of events with a variable number of particles.
The assignment is defined by an injective, non-surjective function informed by physics priors: charged particles are assigned to their originating tracks, while neutral particles are assigned to the cluster where they contribute most of their energy.

We adopt a particle-level objective function following the MLPF approach from Refs.~\cite{Pata:2021oez,Pata:2022wam,Mokhtar:2023fzl}, composed of four loss terms: a particle existence binary classification loss, a multiclass PID classification loss, a binary classification loss for the PU label and a regression loss for the particle four-momentum.
The model is trained to minimize a sum of per-particle loss terms over all input tracks and clusters:
\begin{equation}
    \label{eq:02_mlpf_cms_eq_loss}
    L(Y,\hat{Y}) = \sum_{i} L_\mathrm{cls-binary}(y_i, \hat{y_i}) + L_\mathrm{cls-PID}(y_i, \hat{y_i}) + L_\mathrm{cls-PU}(\alpha_i, \hat{\alpha_i})  + L_\mathrm{reg}(y_i, \hat{y_i}).
\end{equation}

The particle existence binary classification loss $L_\text{cls-binary}$ predicts whether a track or cluster corresponds to a reconstructed particle and is implemented using binary cross entropy.
The multiclass PID loss $L_\text{cls-PID}$ discriminates among particle types (e.g., electrons vs. photons) and is computed using focal loss~\cite{lin2017focal} to address class imbalance by emphasizing harder-to-classify examples. 
The binary classification loss $L_\text{cls-PU}$, implemented using binary cross entropy, identifies reconstructed particles that arose from PU vertices.
The regression loss $L_\mathrm{reg}$ estimates the components of the four-momentum using the mean squared error between the predicted and target values.

To improve performance at high-$\pt$, we weight the per-particle loss terms with a $\sqrt{\pt}$ factor for the \pt and $E$ components, placing greater emphasis on particles in the high-$\pt$ regime. 
We have observed that this weight term improves the jet energy scale and resolution performance over the baseline.

The model is trained end-to-end for eight epochs on a single NVIDIA A100 80GB GPU, requiring approximately 260 hours.
All loss components---binary classification, PID, PU ID, and regression---converge smoothly with no signs of overtraining. 
We observe no performance gains from additional training.
The final model contains approximately 4 million learnable parameters.

\subsection{Input features}
\label{sec:inp}

The input features for each track include the track transverse momentum (\pt) and its uncertainty, the pseudorapidity $\eta$ and azimuthal angle $\phi$ of the particle momentum and their uncertainties (both at the reference point, and extrapolated to the ECAL and HCAL), the Cartesian momentum components, the electric charge, the number of reconstructed hits in the track, and the track vertex coordinates $(v_x, v_y, v_z)$.
The polar angle $\theta$, computed with respect to the beam axis, and $\lambda$, defined as $\lambda = \frac{\pi}{2}-\theta$~\cite{CMS:2014pgm}, along with their uncertainties, are also included.

For calorimeter clusters, the input features include the transverse energy ($E_\mathrm{T}$), $\eta$, $\phi$, electromagnetic calorimeter energy ($E_{\mathrm{ECAL}}$), hadronic calorimeter energy ($E_{\mathrm{HCAL}}$), corrected energy and its associated uncertainty, time and its uncertainty, Cartesian spatial coordinates ($x$, $y$, $z$).
Additional features include the number of hits in the cluster and the cluster size, defined as the weighted standard deviation of the hit positions in $x$, $y$, and $z$.
We also compute depth-dependent cluster shapes by evaluating the weighted standard deviations in $\eta$ and $\phi$ as a function of calorimeter layer (hit depth).

\section{Physics performance in simulation}

\begin{figure}[ht]
    \centering
    \includegraphics[width=0.31\linewidth]{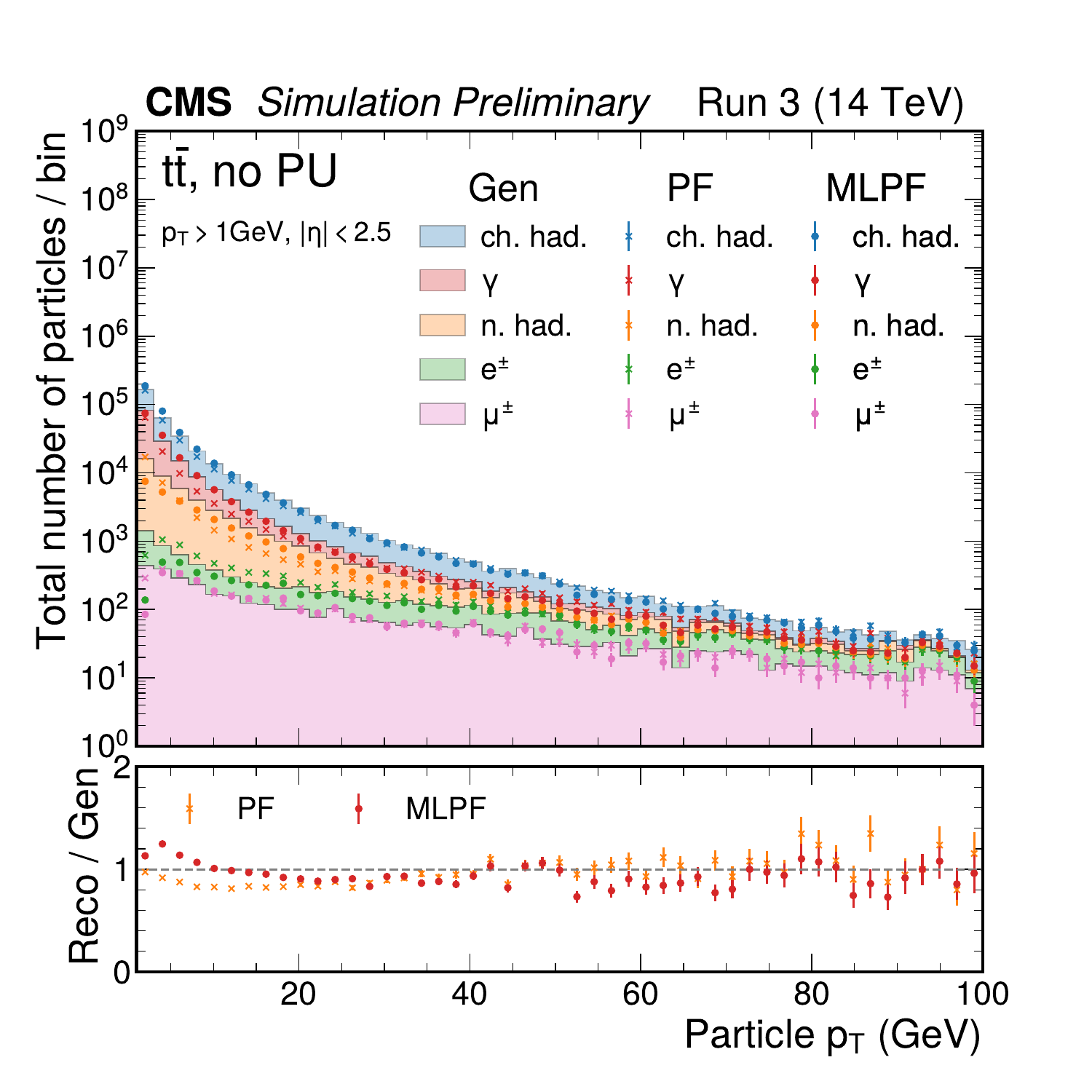}
    \includegraphics[width=0.31\linewidth]{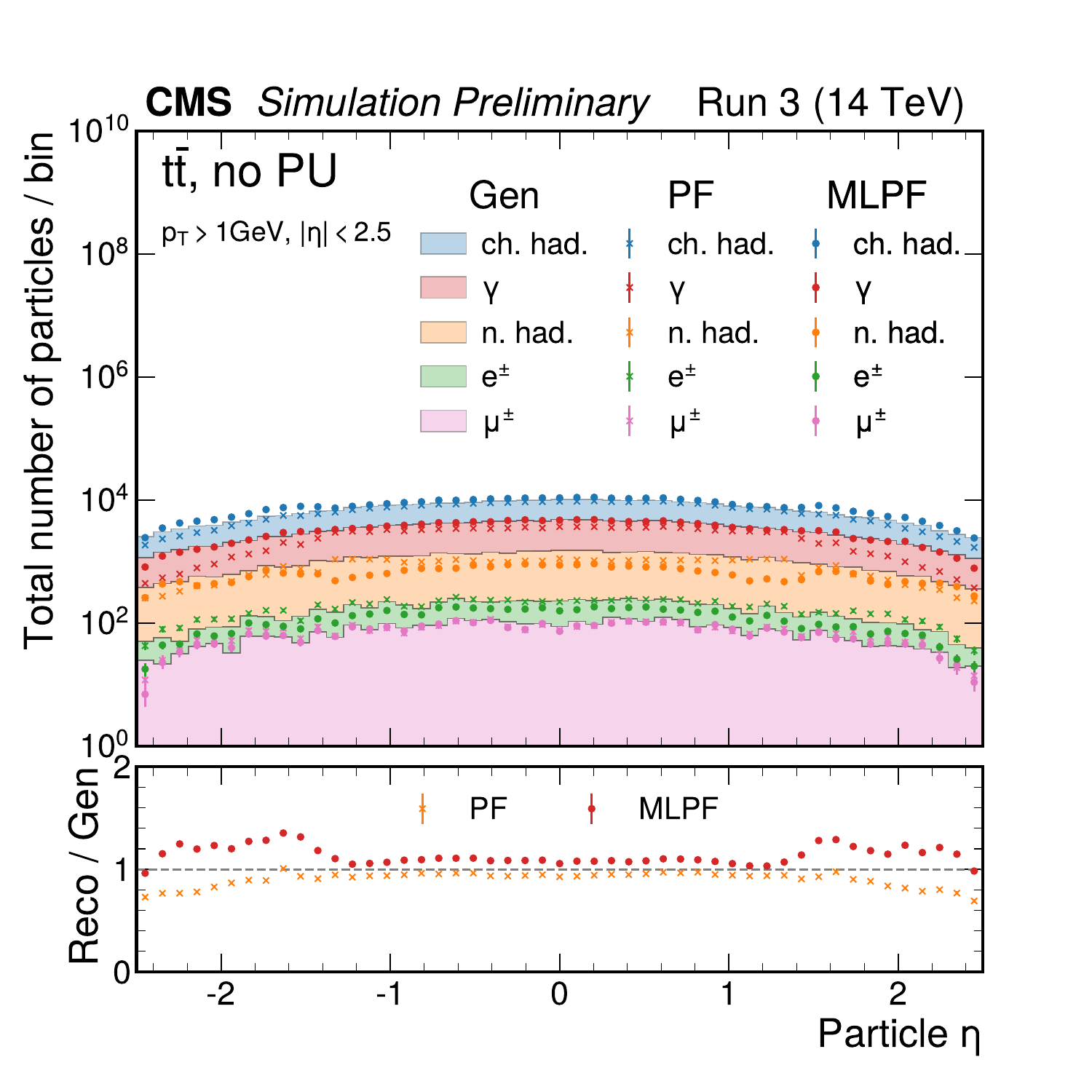}
    \includegraphics[width=0.36\linewidth]{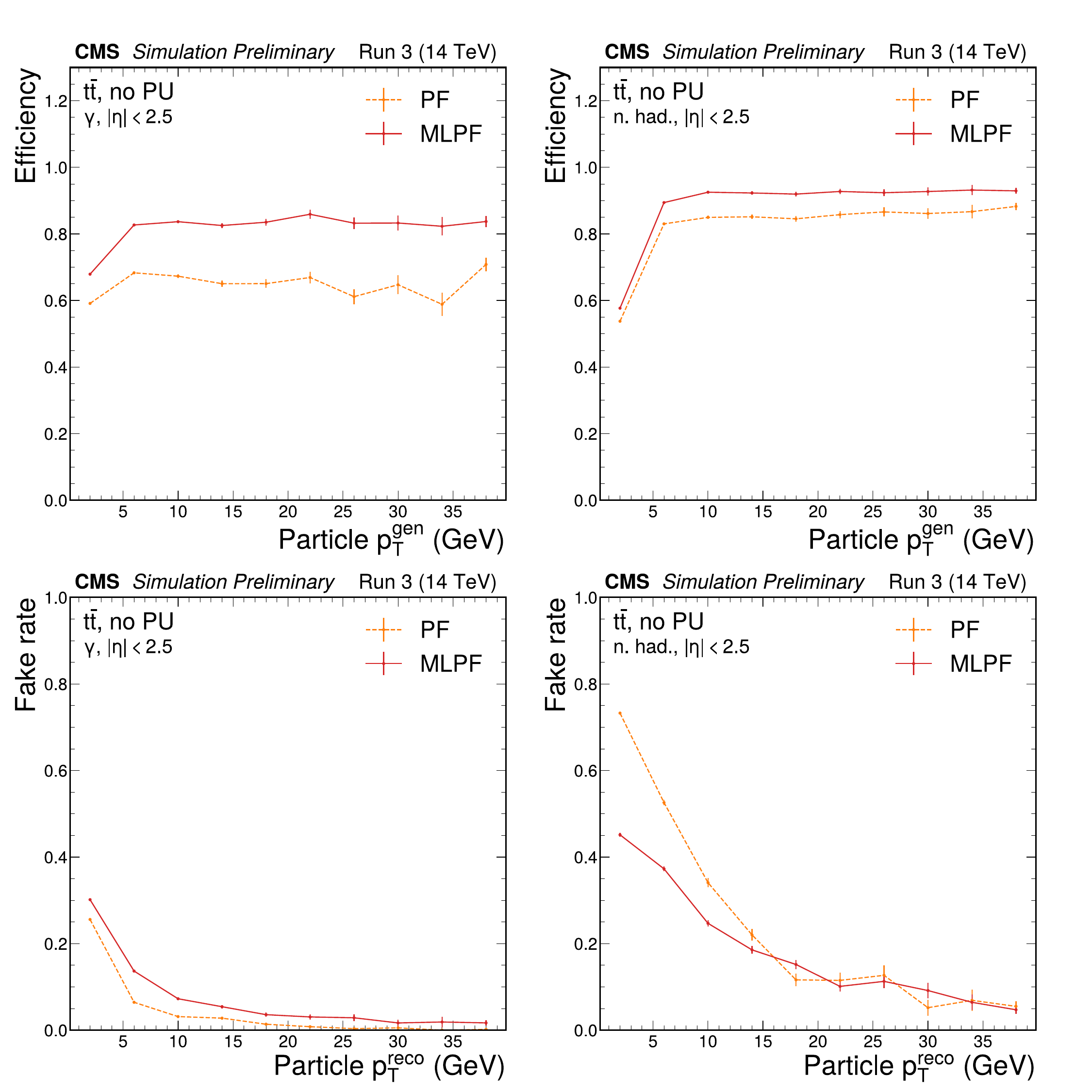}
    \caption{Single particle performance in \ttbar samples.
    Left/center: particle \pt and $\eta$ distributions.
    Right: efficiency and fake rate as a function of \pt for photons and neutral hadrons.
    }
    \label{fig:perf_particles}
\end{figure}

MLPF is evaluated in simulation and compared to PF using \ttbar, and \qcd samples, and shows realistic particle-level performance, with improved reconstruction efficiency for neutral hadrons while maintaining the same fake rate (Figs.~\ref{fig:perf_particles}).

\begin{figure}[ht]
    \centering
    \includegraphics[width=0.32\linewidth]{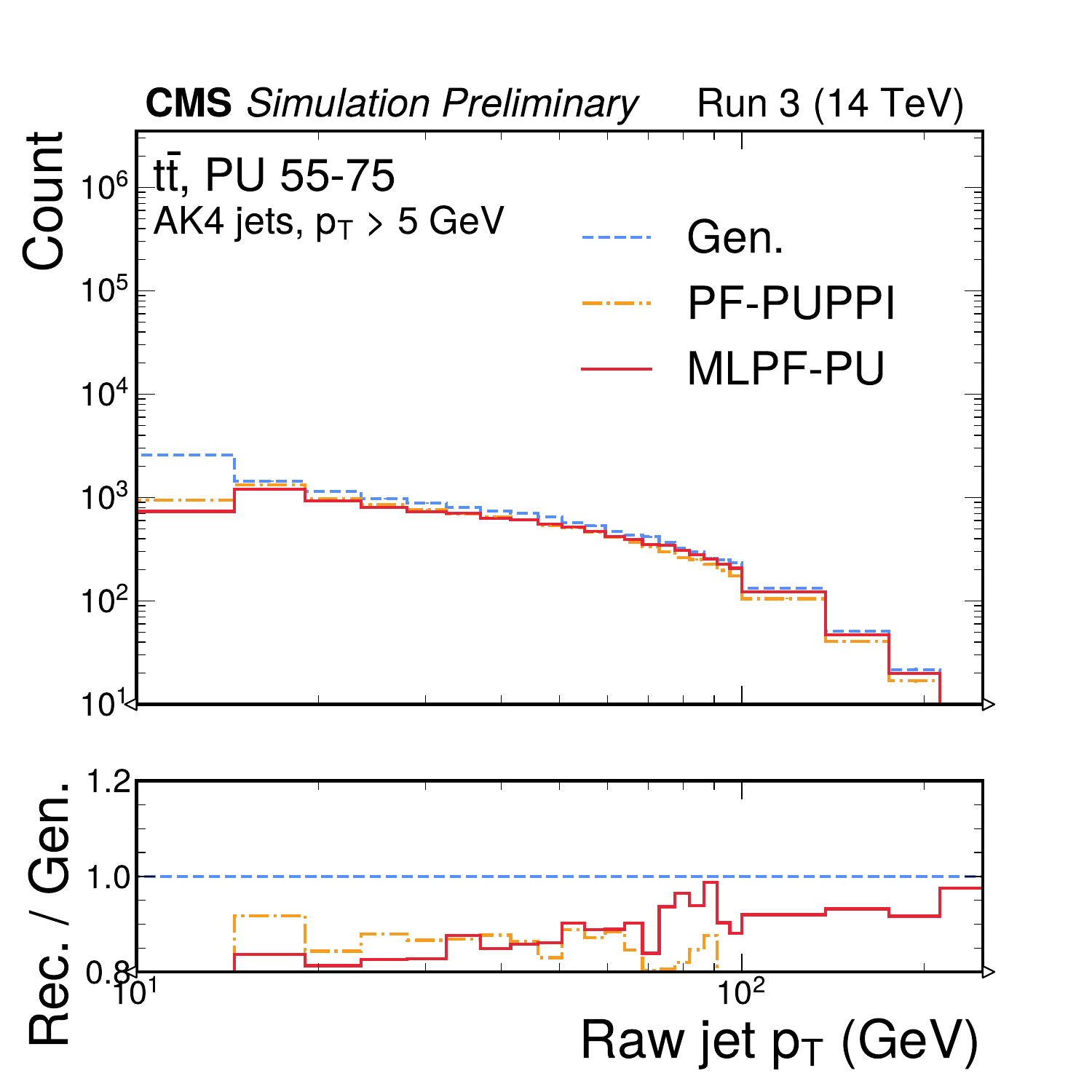}
    \includegraphics[width=0.32\linewidth]{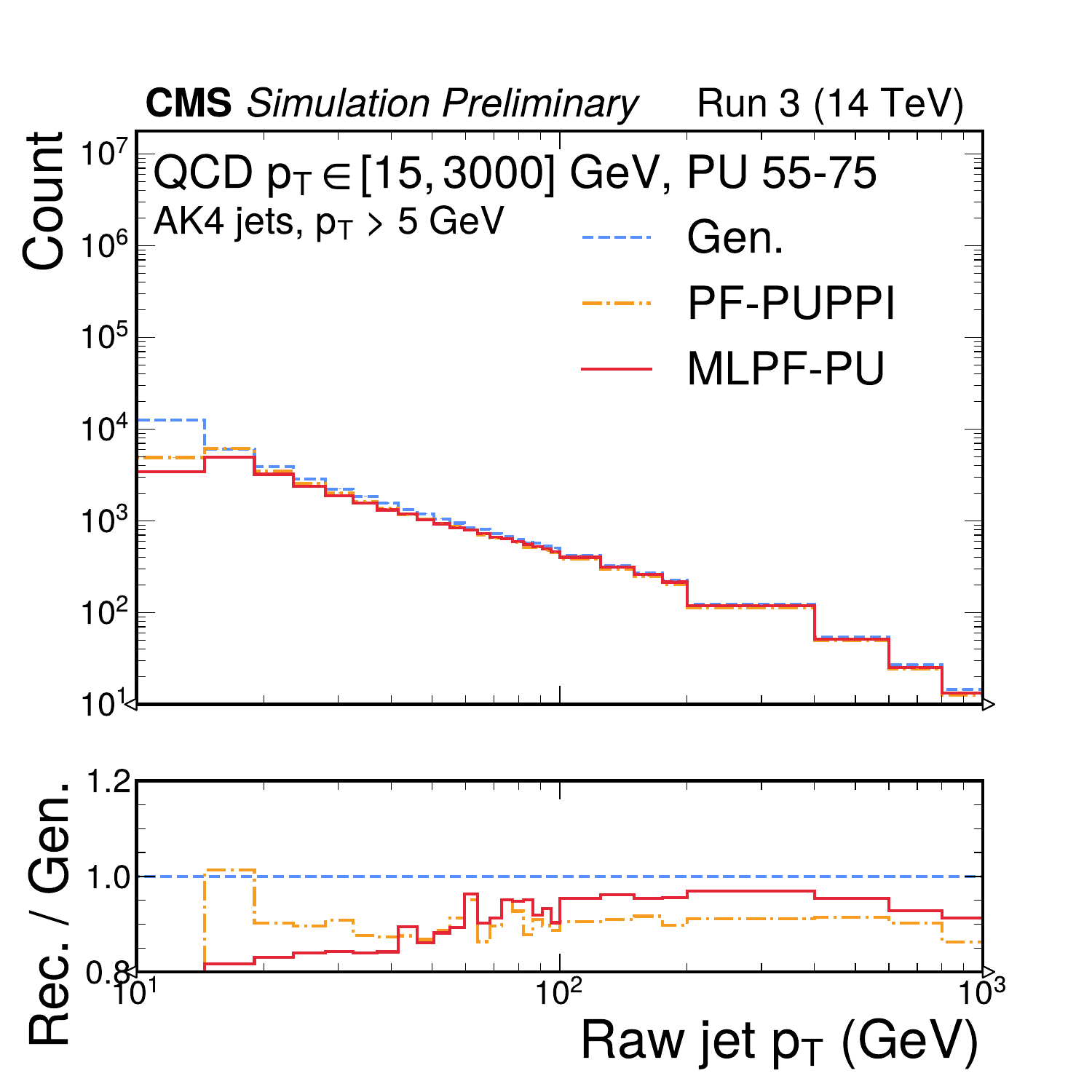}
    \includegraphics[width=0.32\linewidth]{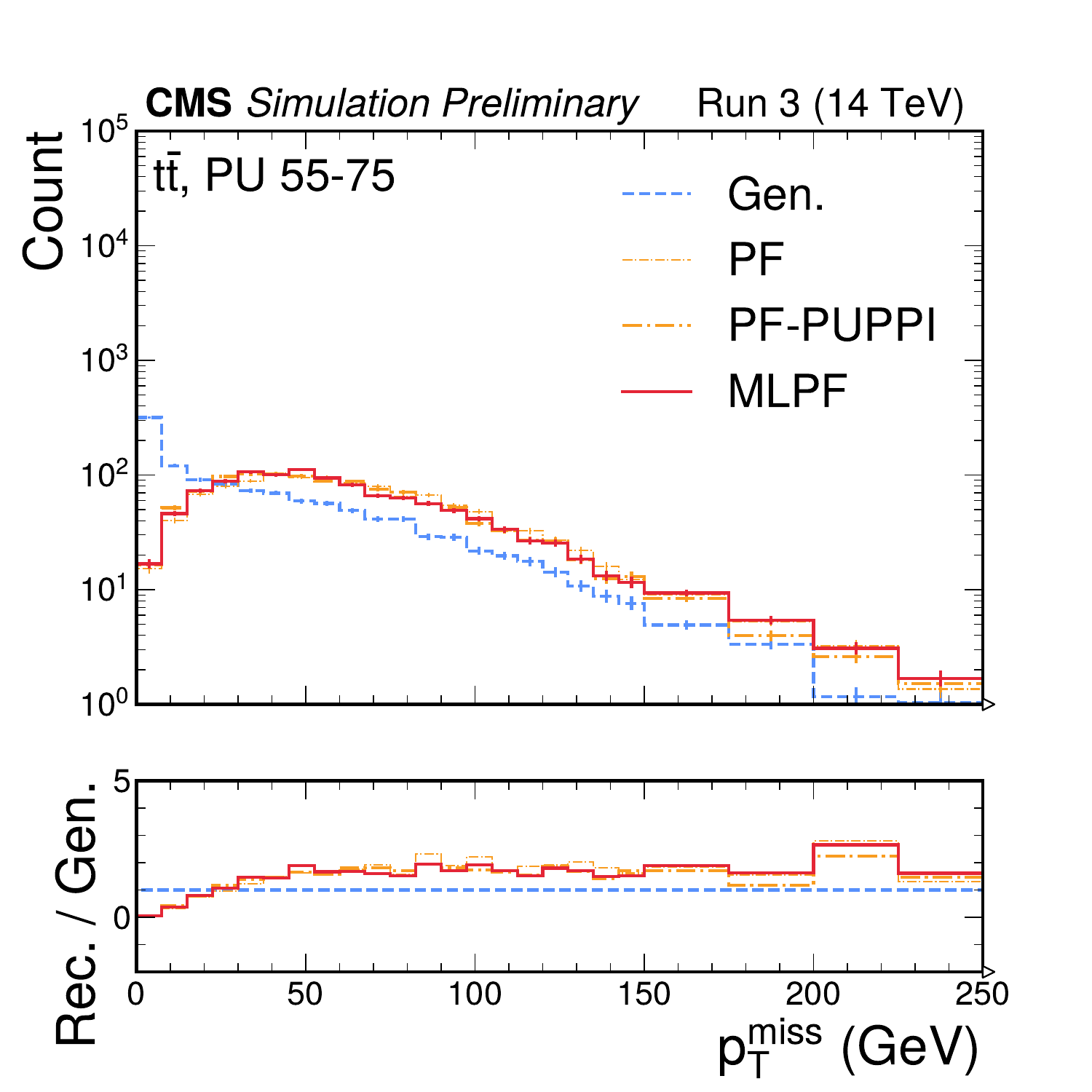}    
    \\
    \includegraphics[width=0.32\linewidth]{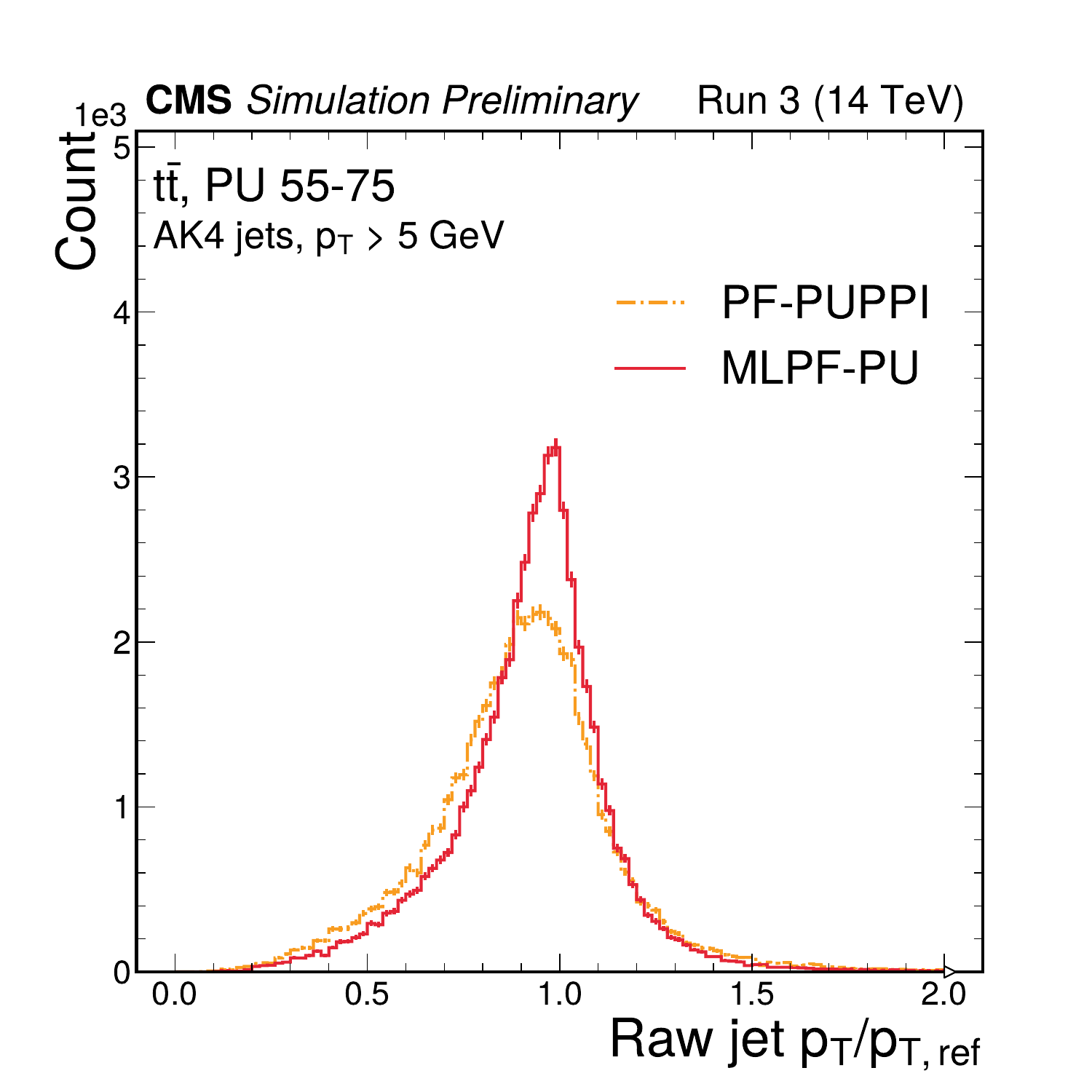}
    \includegraphics[width=0.32\linewidth]{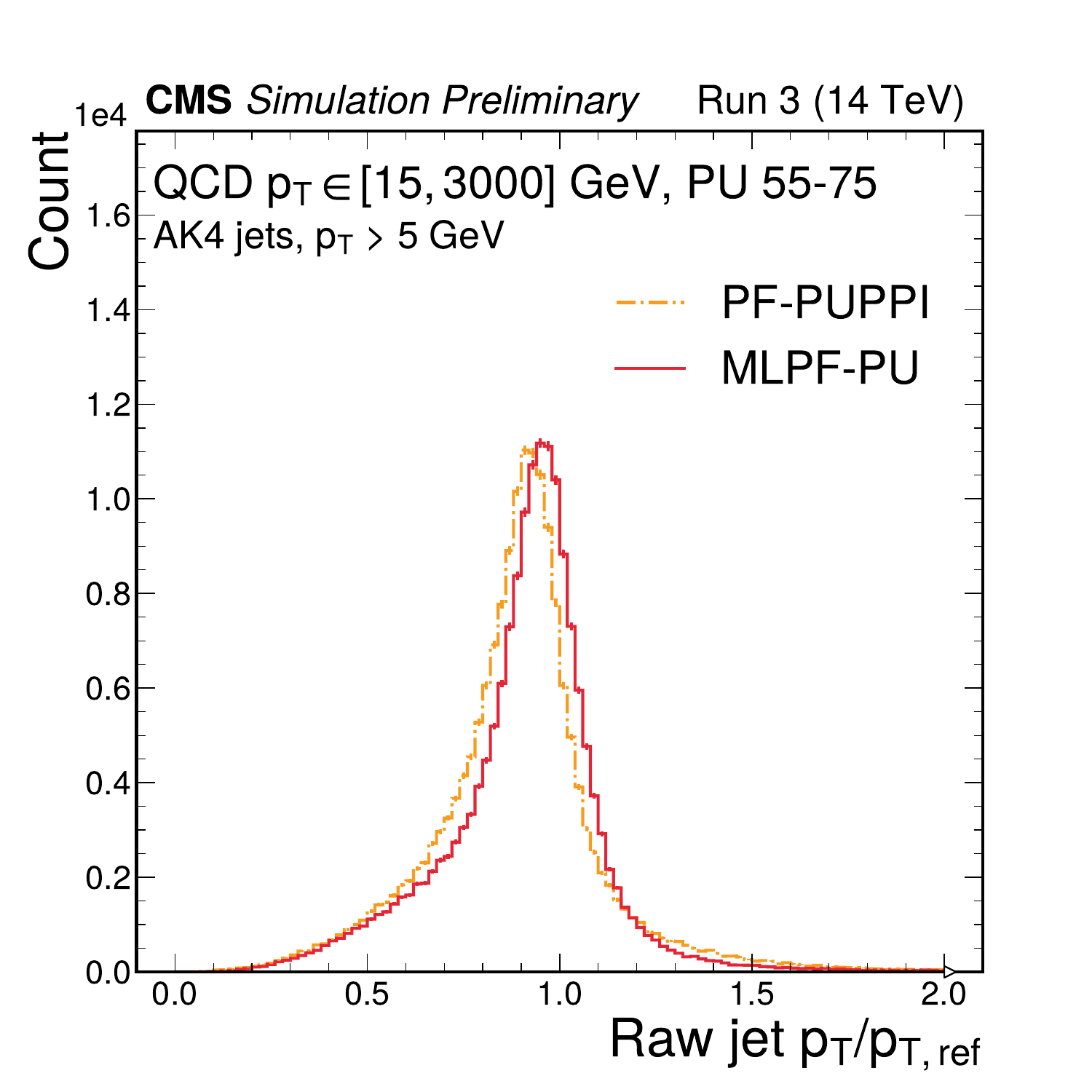}
    \includegraphics[width=0.32\linewidth]{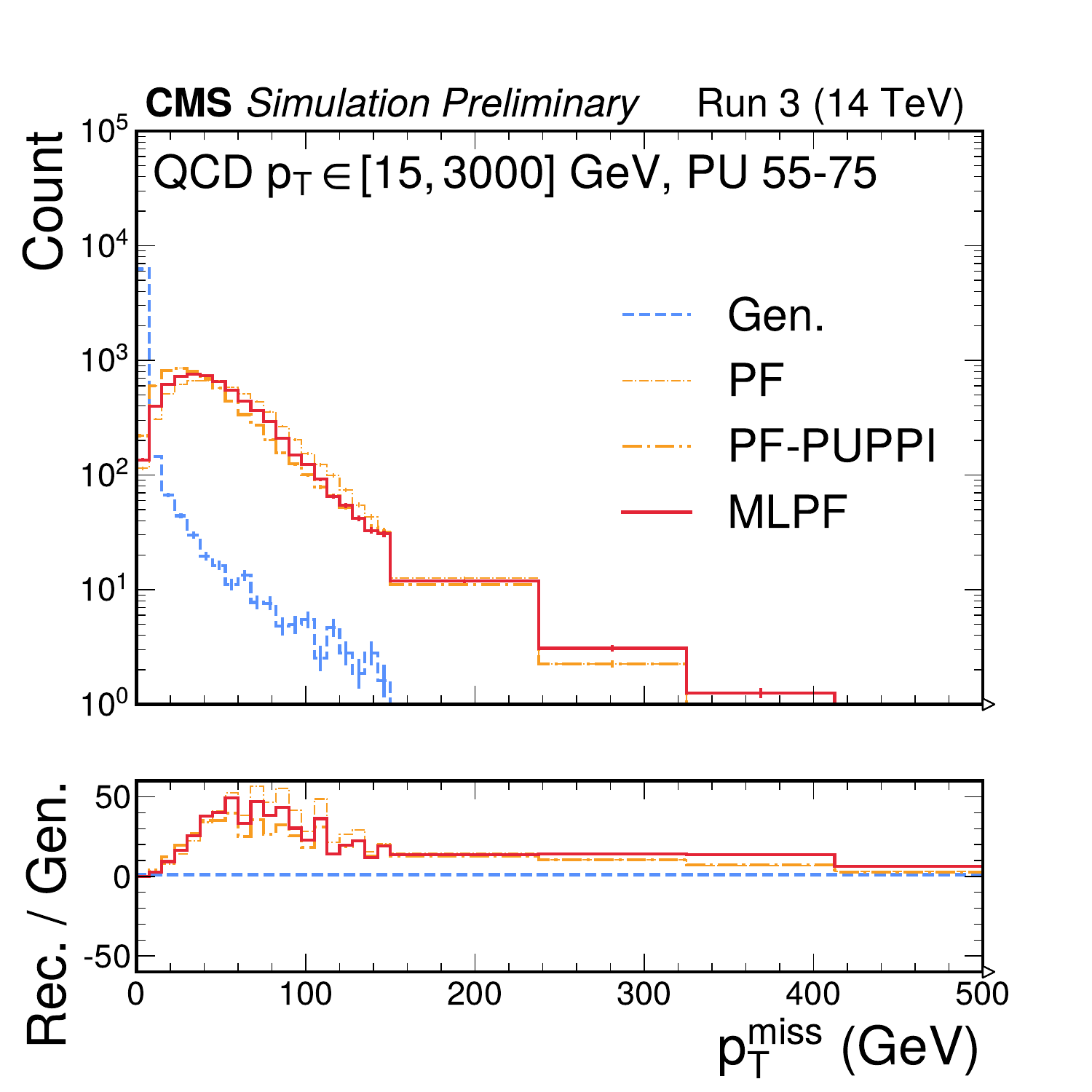}\\    
    \caption{Jet and \met performance in \ttbar and \qcd samples with PU.
    PF uses PUPPI for PU mitigation; MLPF uses per-particle PU scores.
    We show uncorrected jet \pt resolution
    }
    \label{fig:perf_jets_mc}
\end{figure}

Jets are clustered using the anti-\kt algorithm with distance parameters $R=0.4$ (AK4), using either PF or MLPF candidates within \CMSSW.
Jet performance in simulation is evaluated by matching reconstructed jets to generator-level jets within a cone of $\Delta R < 0.4$.
PF jets use the PUPPI~\cite{Sirunyan:2020foa,Bertolini:2014bba} algorithm for PU mitigation prior to clustering. 
MLPF replaces PUPPI with a learned per-particle PU score for neutral and non-tracker charged particles, while using the standard PUPPI treatment for charged particles within the tracker.
Jet and \met performance are comparable to PF (Figs.~\ref{fig:perf_jets_mc}), despite not being explicitly optimized for during training.

\section{Commissioning on data}

\begin{figure}[ht]
    \centering
    \includegraphics[width=0.49\linewidth]{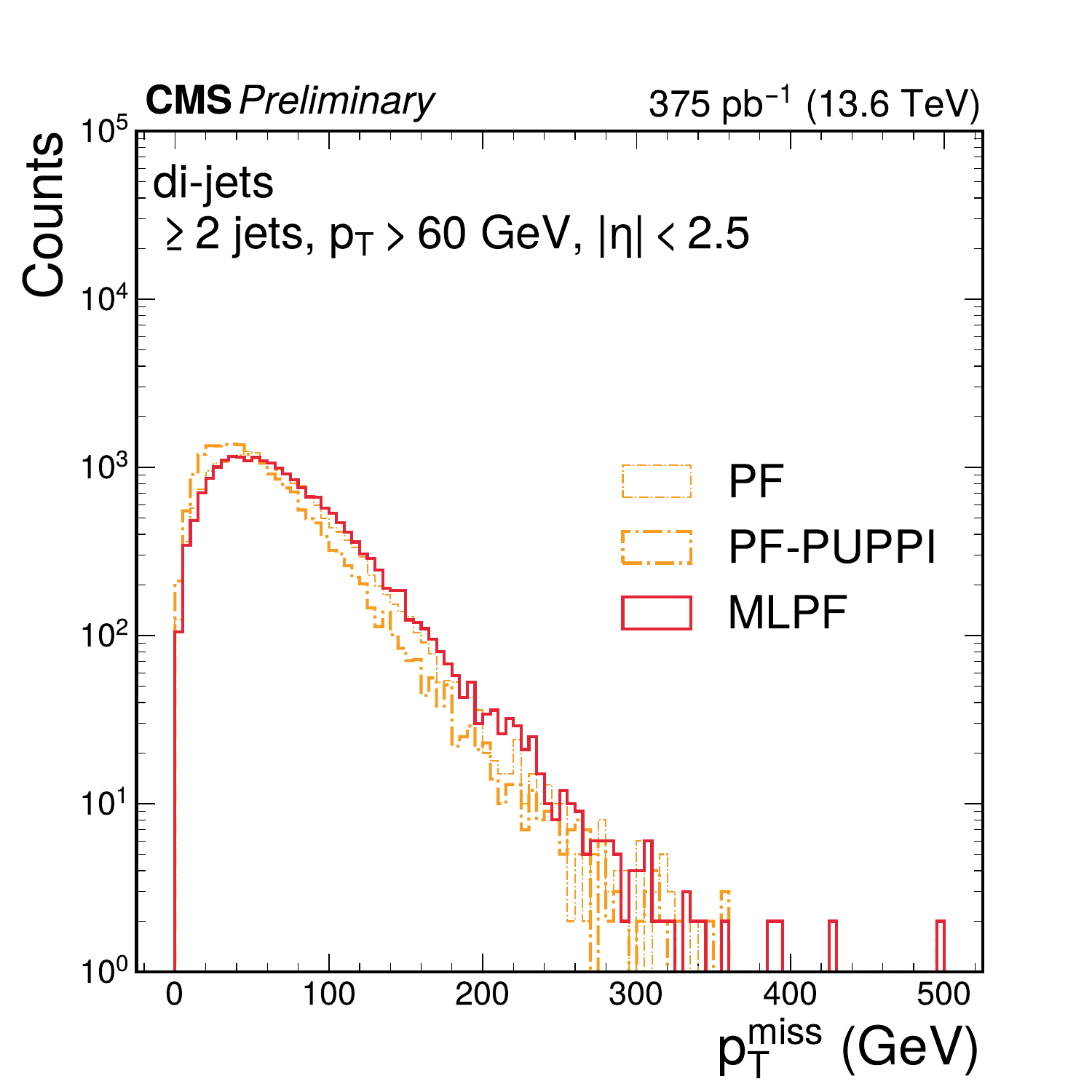}
    \includegraphics[width=0.49\linewidth]{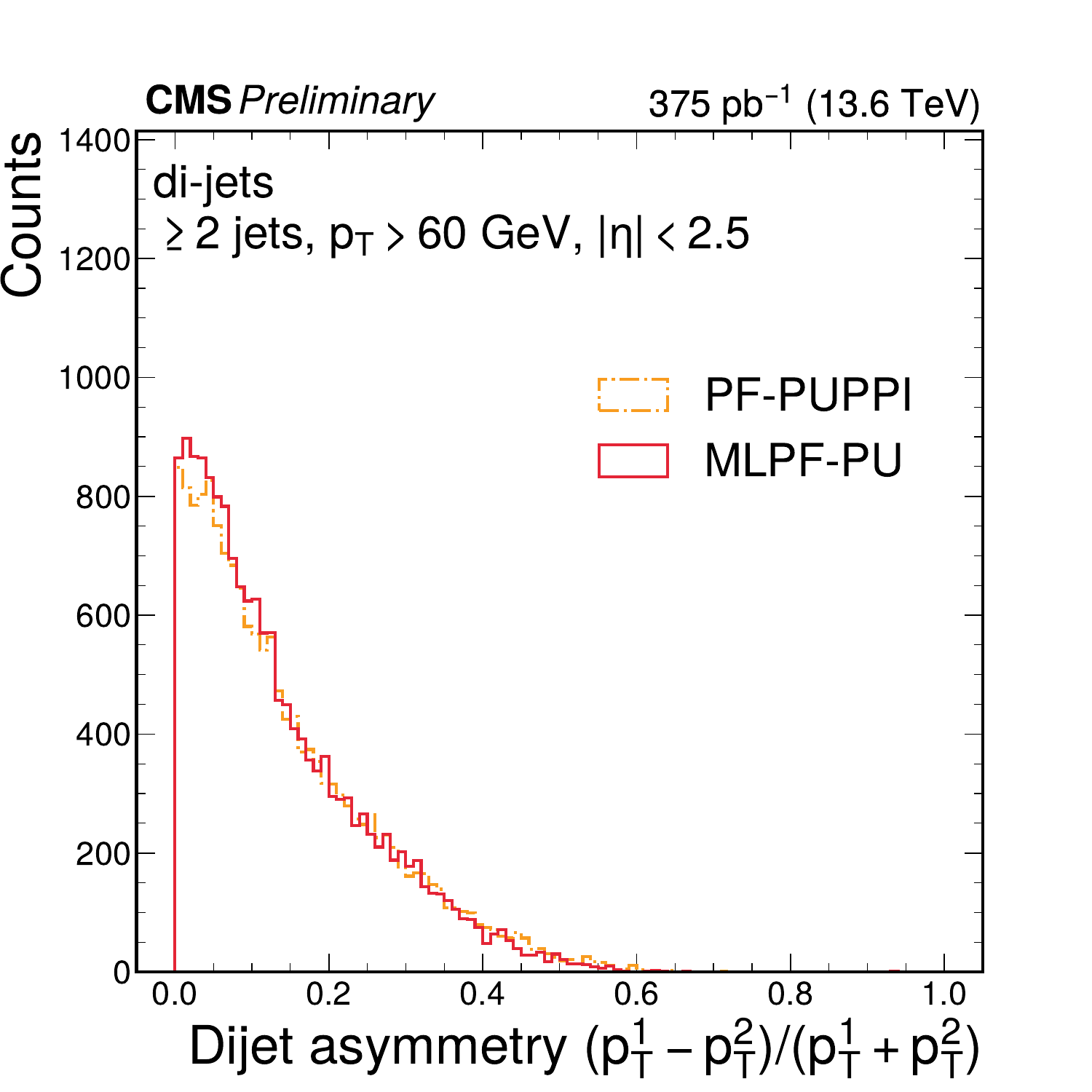}
    \caption{
    Performance on a sample of dijet data collected in Run 3, measure in terms of uncorrected \ptmiss and dijet asymmetry distributions comparing PF and MLPF.
    }
    \label{fig:perfdata}
\end{figure}

MLPF was commissioned on a small sample of data collected during the 2024 data-taking period of Run 3.
The data are required to pass the dijet high-level trigger, and a dijet offline selection.
Reconstructed distributions (e.g., \ptmiss, dijet asymmetry) show agreement between PF and MLPF (Fig.~\ref{fig:perfdata} left), validating the model’s performance on data collected by CMS.

\section{Integration in CMS reconstruction}
\label{sec:cmssw}

\begin{figure}[ht]
    \centering
    \includegraphics[width=0.42\linewidth]{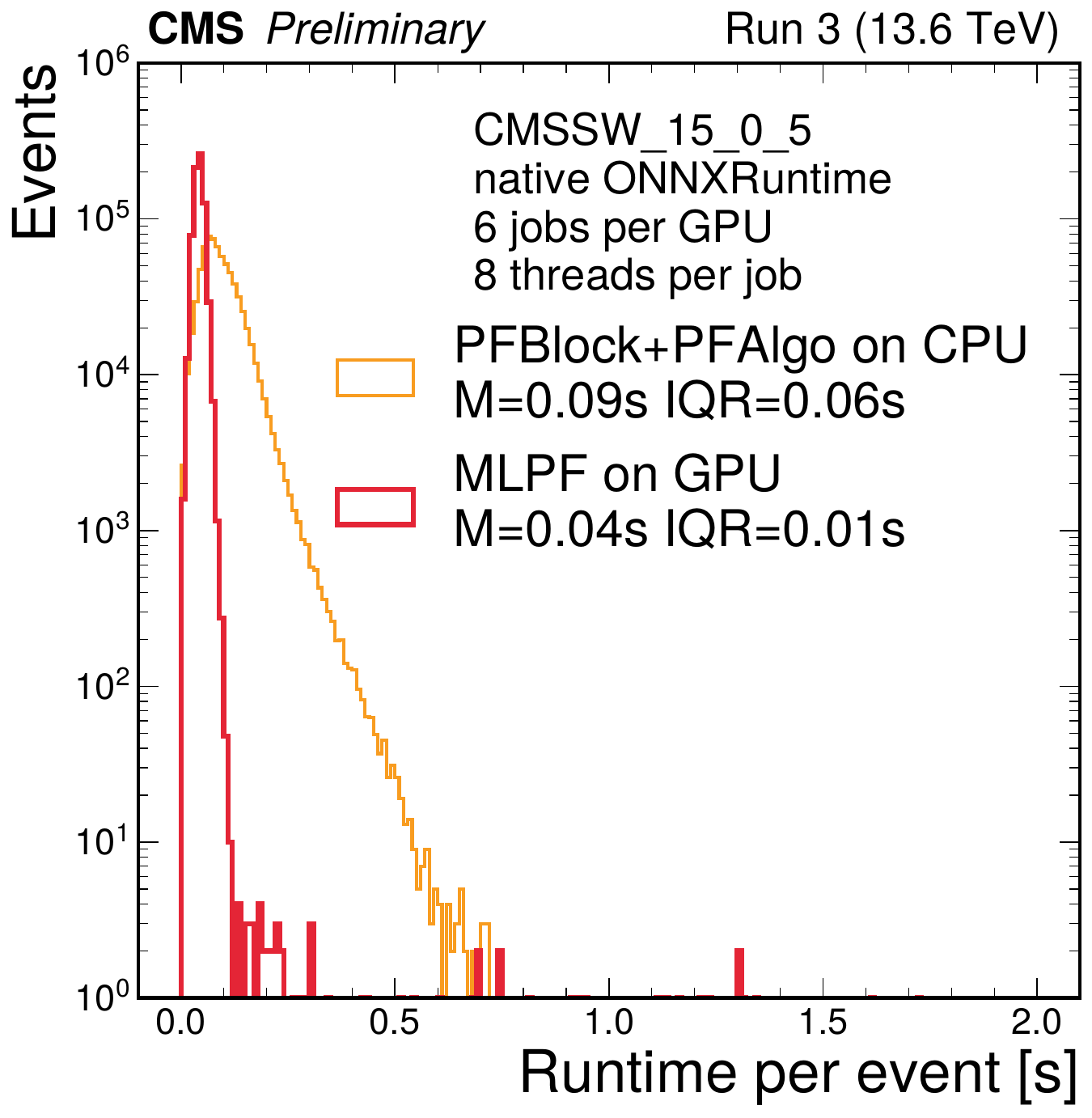}
    \hspace{0.6cm}
    \includegraphics[width=0.42\linewidth]{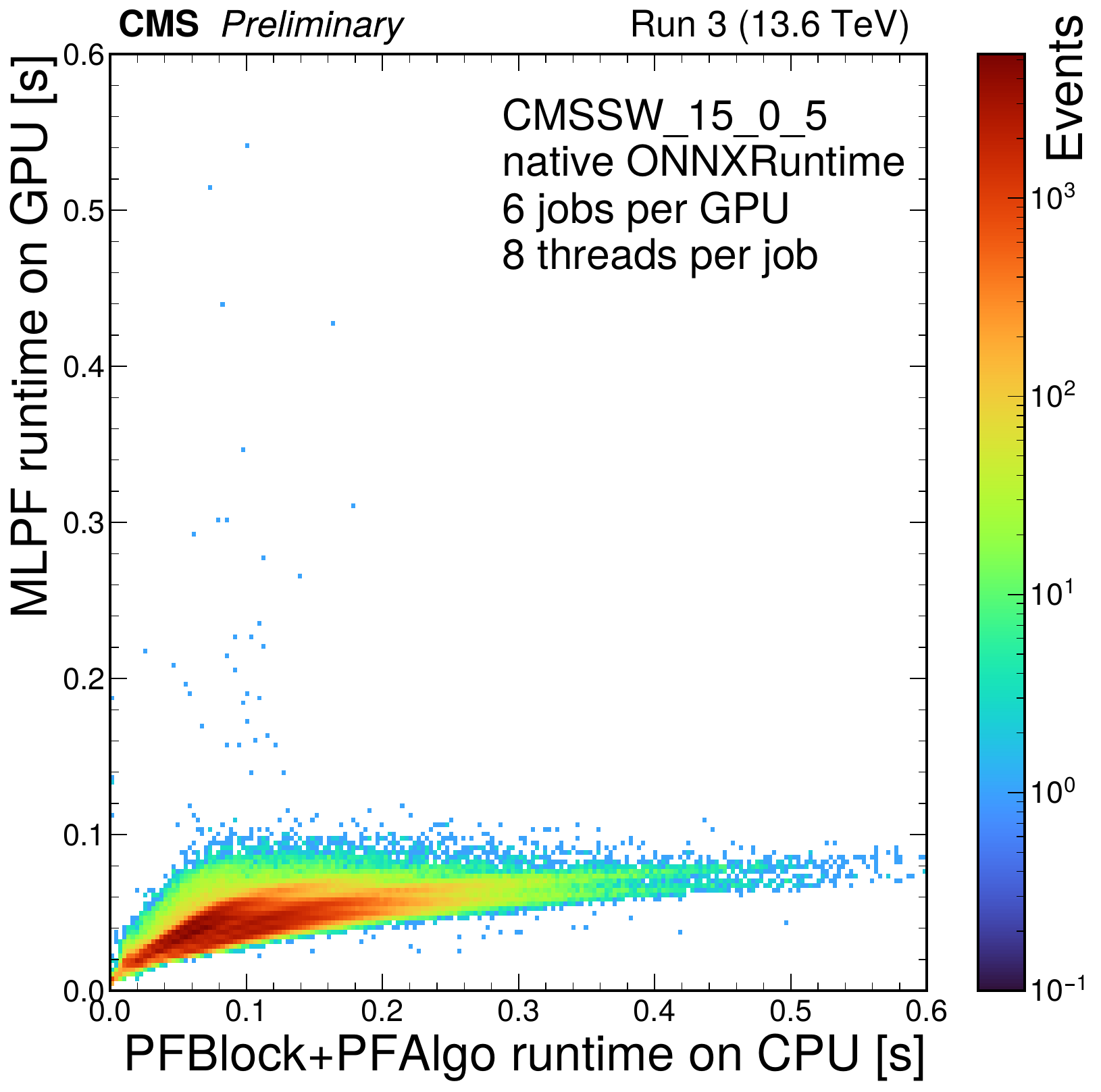}
    \caption{
    Event-level runtime of PF (CPU) vs.\ MLPF (GPU) measured in \CMSSW.
    }
    \label{fig:runtime}
\end{figure}

The model is exported to \ONNX and integrated into \CMSSW via \ONNXRUNTIME with FlashAttention kernels for GPU inference. 
MLPF replaces only the PF reconstruction step, allowing downstream reconstruction modules to run without modification using the particle candidates output by MLPF.
Runtime was benchmarked using six parallel \CMSSW jobs on a dedicated test machine, each using 8 CPU threads and 1/7th of an NVIDIA A100 80GB GPU.
MLPF achieves a $\sim$2× speedup over PF (Fig.~\ref{fig:runtime} right), with runtime of $\sim$40\,ms/event on GPU vs.\ 90\,ms/event for PF on CPU.
Model outputs are stable across hardware and repeat runs.

\section{Summary}
\label{sec:sum}

We present a machine-learned algorithm for particle-flow (MLPF) reconstruction in the CMS experiment, based on transformers with FlashAttention, demonstrating comparable physics performance, including PU mitigation, while achieving a factor of two reduction in runtime. 
The model is integrated in \CMSSW and validated on simulation and data.
Future work includes inference optimization and deployment under HL-LHC Phase-2 conditions.

\bibliographystyle{cms_unsrt}
\bibliography{references}


\end{document}